\documentclass[%
 reprint,
 amsmath,amssymb,
 aps,
floatfix]{revtex4-2}
\usepackage{physics}
\usepackage{graphicx}
\usepackage{dcolumn}
\usepackage{bm}

\begin{document}

\preprint{APS/123-QED}

\title{Multipartite high-dimensional entangled state generation through soliton-induced dynamical Casimir effect on a chip}

\author{Ali Eshaghian Dorche}
\email{a.eshaqian@gmail.com}

\author{Ali Adibi}%
 \email{ali.adibi@ece.gatech.edu}
\affiliation{%
 School of Electrical and Computer Engineering, Georgia Institute of Technology, $778$ Atlantic Drive,
Atlanta, Georgia $30332$, USA
}%

\date{\today}

\begin{abstract}
An integrated photonic approach for complex quantum state generation through dynamical Casimir effect (DCE) is demonstrated. This approach provides a scheme to realize multipartite high-dimensional entangled states in the microwave (MW) and terahertz (THz) regimes, through the modulation in a MW-resonator coupled to an optical microresonator supporting temporal optical solitons. The states at the MW-resonator evolve from the ground state, realizing real-photons from the virtual photons at the ground state. The periodic modulation of the MW-resonator through a Kerr-induced refractive index change in the optical microresonator, along with the localized spatial distribution of the dissipative Kerr solitons (DKSs), enables photon-pair generation and inter-mode coupling at the MW-resonator. This allows generating highly persistent multipartite high-dimensional entangled states in a wide range of spectrum. The proposed approach paves the way for a hybrid integrated platform for generation of multipartite entangled qudits (high-dimensional qubits) at the MW and THz regimes using highly coherent ultra-short optical pulses in a monolithic integrated platform. This architecture can act as an entangled state source, as a necessary resource for exploiting a wide range of quantum protocols from fault-tolerant computing to enhanced sensing and teleportation.
\end{abstract}

\maketitle


\section{\label{sec:level1}Introduction}

Quantum mechanics, thanks to its unique and classically counter-intuitive features, e.g., superposition and entanglement of states \cite{1}, has received extensive attention during last two decades for its capabilities in handling naturally hard problems \cite{2}, enhanced sensing beyond the classical limits \cite{3, 4}, and new protocols of communication (e.g., teleportation) \cite{5}. Another important feature of quantum mechanics, known as no-cloning theorem \cite{6}, form the basis for quantum key distribution (QKD) \cite{7}. Entangled states are among the essential resources for implementing these applications \cite{8}. For example, teleportation of a quantum state relies on sharing an entangled state between different parties to implement a measurement, exchange the measurement outcome through classical communication means, and finally apply the appropriate quantum logic to transfer the quantum state. During this process, the entangled state is used as a resource to realize the teleportation protocol upon local operations and classical communication (LOCC).

Entangled states also form the basis of fault-tolerant quantum computing \cite{9, 10, 11, 12, 13}. Considering the limited coherence time and fragility of qubits, encoding information on multi-partite entangled quantum states is indispensable for error-correction protocols. This becomes more important for superconducting qubits and other quantum protocols in the MW spectrum where the coupling to the thermal reservoir severely suppresses the coherence time of qubits \cite{14}. Considering versatile applications of quantum mechanics in a wide range of spectrum, from MW to ultraviolet (UV), for information processing, sensing, communication, and spectroscopy, developing capabilities to generate entangled states at a desired wavelength is of practical importance \cite{15, 16, 17, 18, 19, 20}.
   
Entangled photon generation has been the subject of intense research and developments for a long time, starting from free-space setups, mainly as a testbed for the study and evaluation of the hypotheses of quantum mechanics \cite{21}, to more integrated structures both at MW and optical regime to implement quantum devices and systems \cite{14, 22}. Within the optical  spectrum, frequency-bin entangled states generated through spontaneous four-wave mixing (SFWM) in a high-quality-factor (high-Q) microresonator \cite{23, 24, 25, 26, 27}, or spontaneous parametric down-conversion (SPDC) in periodically polled lithium niobate (PPLN) \cite{28, 29, 30} are the most adapted configurations, providing a large number of heralded photons forming the entangled states. However, these states suffer from low persistency, i.e., once subject to measurement of  photons in a mode, the state collapses into a fully separable one \cite{26}. Recently, cluster state formation through post processing of hyper-entangled states, generated through driving a high-Q optical microresonator by time-bins, has shown promises toward higher-dimensional entangled states. However, this comes at the expense of complex post-processing units to coherently access and manipulate terms in the hyper-entangled state, using a synchronized phase modulator and a fiber Bragg array in a self-referenced and phase-stable loop \cite{27}. This highlights the challenges toward fully integrated solutions for further extending the entangled state dimensions to enhance the quantum computation capabilities for practical purposes. This protocol is also well suited for operation in the optical spectrum, forming low-loss and long-distance connections between different nodes of a quantum network. However, local connections between the physical qubits in these nodes require microwave photons. This is mainly because the energy-gap in superconducting qubits and energy splitting in the ground state of atoms/ions lies in the microwave regime (i.e., a few GHz). Thus, connecting, transferring, and entangling stationary qubits (i.e., SC-qubits, atoms, ions) requires entangled microwave photon states. Among different platforms, SC based systems received attention as a scalable solid-state platform enabling flexible, macroscopic objects with strong coupling to external fields. This is in contrary to the atom/ion based processors, which suffer from either low scalability, weak coupling to external field (solid state vacancies), or not being scalable solid state solution (trapped atoms/ions). On the other hand, atom-based qubits leverage their long coherence time which is superior to that of SC qubits. Therefore, it seems not a single solution addresses all the requirements for an efficient quantum processor. A fully integrated quantum network would combine the advantages of both realms, thus a hybrid platform containing SCs, trapped atoms/ions, and photonic devices. Furthermore, beside the quantum processing, there are significant applications at the MW, THz and infrared (IR) regimes, for biomedical imaging \cite{16}, spectroscopy \cite{17}, and communication \cite{22}, to benefit from the developed architectures providing entangled states at these frequencies.

Generating entangled states at low frequencies in the MW and THz regimes is more challenging due to the stronger coupling of qubits to the thermal reservoir \cite{14}. SC artificial-atom entanglement has been proposed through the dynamical Casimir effect (DCE) by modulating a tunable inductance, in the form of a SC quantum interference device (SQUID), connected to one end of SC-resonators, containing the SC-qubit \cite{31}. Under resonant conditions, this generates photon-pairs shared between SC-qubits to entangle them \cite{31}. This approach has the potential to generate genuine multipartite entanglement (GME), e.g., a three-particle entangled state. This process depends on the fast modulation of the magnetic flux, which limits its applicability to low frequencies below the plasma frequency of the SQUID, on the order of a few tens of GHz. Thus, such proposals have been limited to DCE-induced photons at the low-frequency range of MW spectrum in a single resonant mode \cite{31}. With emerging applications in the MW regime (e.g., 5G communications, internet of things (IoT), etc.), and intrinsic signatures of molecules in the MW-THz and IR spectral regions, developing a technology for generating entangled photon states with high degree of entanglement and persistency in these spectral regions are of significant importance. In this paper, we demonstrate a new approach for multi-partite high-dimensional entangled state generation with high-degree of entanglement persistency upon measurements. This is achieved through DCE in a MW-resonator coupled to an optical-resonator supporting optical DKSs. This approach enables highly entangled states with a good degree of control on the frequency of the generated photons by adjusting the radius of the optical microresonator, paving the way toward fully integrated structures for high-dimensional entangled state generation, suitable for quantum information processing, quantum communications, and spectroscopy, just to name a few important applications. In Section \ref{sec:level2}, we explain the underlying dynamics of the hybrid optical-MW architecture, leveraging high-speed modulations in the optical microresonator perturbing the MW-resonator. Section \ref{sec:level22} provides the effective Hamiltonian to describe the dynamics of the vacuum states in the MW resonator, which is used to study photon generation in a limited number of MW-resonant modes. Entanglement in the generated state is studied in Section \ref{sec:level3}, and the outlook for implementations are discussed briefly in Section \ref{sec:level4}.

\section{\label{sec:level2} Soliton based DCE on a chip}

Dissipative Kerr solitons generated inside a high-Q optical microresonator consists of a series of coherent equidistant markers in the frequency domain, forming short optical pulses with significant peak powers inside the microresonator \cite{32, 33, 34, 35}. Thanks to the major advances in dispersion engineering and microfabrication techniques, on-chip Kerr-comb solitons with FSRs as low as a few GHz up to hundreds of GHz are demonstrated with reasonable pump 
powers \cite{36, 37, 38}. A well-defined FSR defines the temporal periodicity of the generated soliton. The high peak-power of the soliton inside a high-finesse microresonator results in observable nonlinear effects. 

An optical microresonator driven by a unidirectional or bidirectional coherent CW laser, can support a range of solutions (fixed-points), from a single soliton to soliton crystals and counter-propagating (CP) solitons. While the two first cases are generated through driving a high-Q microresonator with a coherent CW source, the third one is formed through bi-directional pumping of the same optical microresonator. The CP solitons have been of interest recently in the perspective of studying their dynamics and their application in spectroscopy \cite{39, 40, 41}. Among different CP-soliton dynamics, we will focus on locked CP solitons with identical FSRs \cite{39}. Efficient soliton generation requires microresonators with high Qs to reduce the required power for forming the soliton. High-Q optical microresonators have been demonstrated in a wide range of material platforms, e.g., silicon nitride (SiN) \cite{42, 43, 44}, lithium niobate (LiNbO3) \cite{45}, and aluminum nitride (AlN) \cite{46}. SiN is the most adapted platform due its CMOS foundry compatibility, wideband transparency window, and reasonable third-order nonlinearity. Throughout this paper, we consider an optical microresonator formed in SiN, buried in a silicon dioxide (SiO2) clad. 

\begin{figure}[t]
\centering
\includegraphics[trim = 0 0 0 0, width=0.9\linewidth]{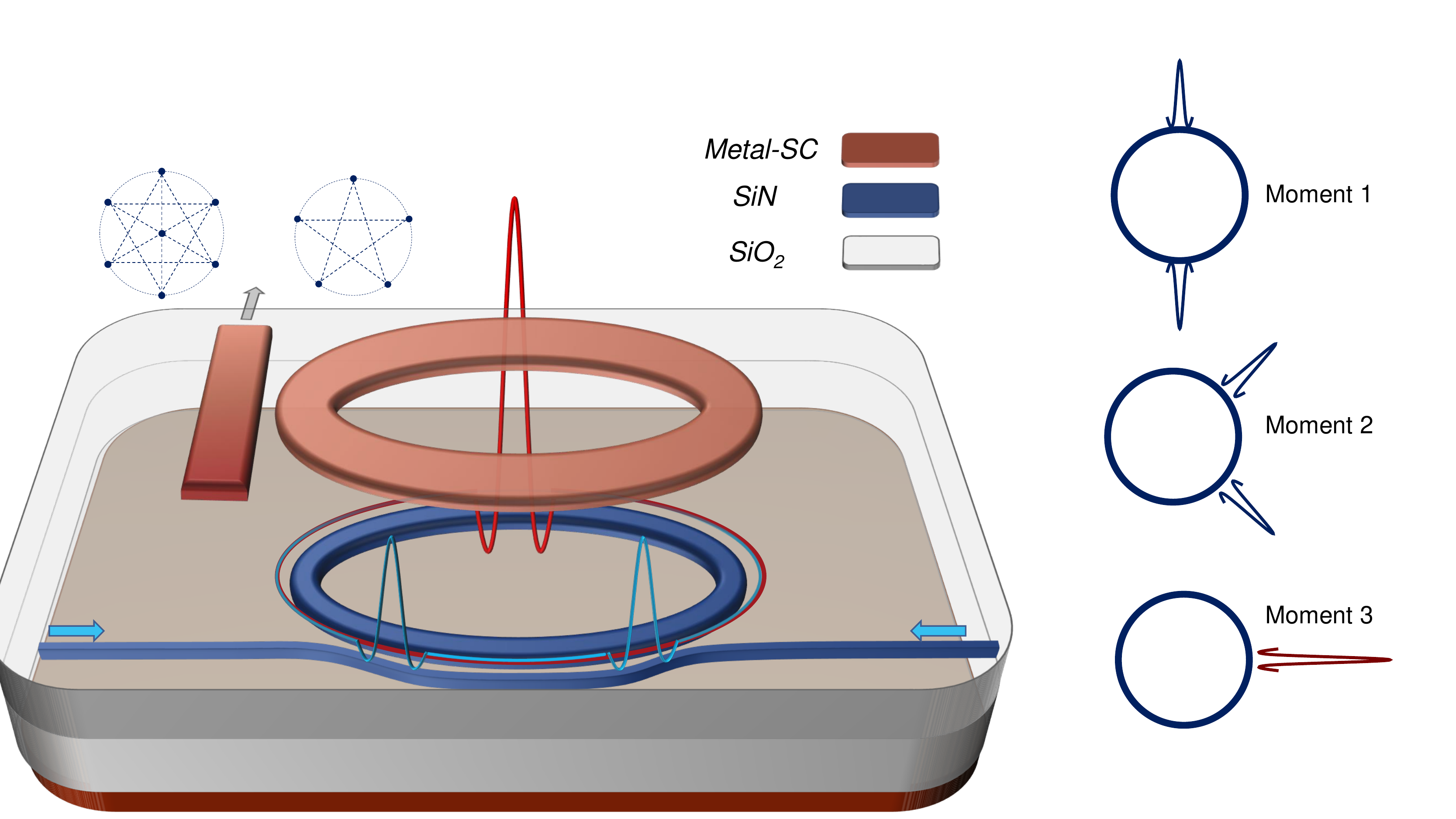}
\caption{\label{fig1} Schematic of the structure for multipartite high-dimensional entangled state generation through DCE in a MW-resonator, coupled to an optical microresonator supporting counter-propagating DKSs. The optical and MW resonators are coupled to bus waveguides in the form of slab, and coplanar strip-line waveguides, respectively. The optical microresonator is formed in a Kerr medium, so CP solitons at different moments result in different refractive index changes inside the optical microresoantor. This would induce small periodic perturbations to the optical and MW resonators. The ground states in the empty MW-resonator experience the periodic perturbation, which under resonant conditions leads to real-photon generation. The inhomogeneous refractive index in the MW-resonator, with spatial distribution following the intensity of the temporal DKSs, allows inter-mode coupling between MW-resonances by exciting and scattering of photons among different modes. Considering wide spectral coverage of DKS, i.e., the large number of comb lines, soliton-induced DCE excites a large number of modes in the MW-resonator, with inter-mode coupling in-between them. This paves the way toward states for quantum error correction, e.g., 5, 7 qubit codes for quantum error correction protocols as shown by the two circles on the top left. The parties are shown by small solid circles with dotted lines show the connection between them.}
\end{figure}

Figure \ref{fig1} represents the schematic of the proposed structure for studying MW-photon generation. The optical microresonator and the bus waveguide are formed in the SiN layer, and the MW-resonator, in the form of a coplanar ring coupled to a micro-strip, is formed above the optical microresonator in a metallic region, with another metallic layer as a ground plane beneath the optical layer (see Fig. \ref{fig1}). The distance between the metallic layers and the SiN layer are adjusted to ensure the optical field profile (and thus, the Q of the optical microresonator) in SiN is not affected. There are other hybrid MW-optical configurations, in the form of lateral MW and optical resonators, or in a hybrid form with electrodes placed above, and next to the dielectric forming the optical resonator. These structures would allow manipulating field overlap between MW and optical resonators, or in the case of SC MW-resonator allow for better satisfying the expulsion of magnetic field below the critical temperature (Meissner effect) \cite{72}.  Here, without loss of generality, we consider the configuration shown in Fig. \ref{fig1}, and we consider a fixed overlap along the resonators. Further discussions about the fabrication considerations of the MW-resonator using superconductors, the operating frequency, and the state-of-the-art Qs are discussed in Section \ref{sec:level4}. For the rest of this paper, we consider the metallic pads in Fig. \ref{fig1} as perfect conductors (PCs).

A DKS, as a temporal soliton, perturbs the effective refractive index in the optical microresonator in Fig. \ref{fig1}, effect of which is experienced by the MW-resonator through a coupling factor determined by the coupled-mode theory (CMT) \cite{47}. This provides a periodic perturbation of the MW-resonator with the well-controlled periodicity of the DKS ($T_{opt}$), determined by the group index of the optical microresonator and its radius. The FSR in the optical microresonator ($FSR_{opt}$) is a well-controlled parameter, ranging from a few GHz to a few THz, supporting stable and coherent DKSs. Thus, the MW-resonator is modulated with harmonics of $FSR_{opt}$. Considering resonant conditions for generating photon-pairs through DCE in a MW-resonator, the modulation of the MW-resonator should be an integer-multiple of its fundamental resonant frequency, which is the same as its FSR ($FSR_{MW}$). By controlling the radius and configuration of the MW-resonator, it is possible to adjust the properties of its resonant modes (e.g., resonance frequencies or $FSR_{MW}$). We can conceive two different approaches for modulating the MW-resonator: first, using co-propagating DKSs (a single DKS or soliton crystals). A single soliton modulates the MW-resonator with periodicity $T_{opt}$ in a structure where the coupling factor between the MW and optical modes is not constant along the propagation direction. This is required to modulate the MW-resonator by varying instantaneous total optical path length in the MW-resonator that can be achieved by an engineered overlap of the MW and optical modes, which we may call it the masking coefficient (or $M(\theta)$). The second scheme is the use of CP solitons. In this case, even with a constant coupling factor between the optical and the MW resonators along the propagation direction (i.e., $M(\theta)=1, \, \forall \theta$), the MW-resonator is modulated through the interference of the coherent CP solitons in a Kerr nonlinear medium. The CP solitons constructively interfere once fully  overlapped, leading to an enhanced intensity in the nonlinear medium, while at other moments the peak of one soliton experiences the pedestal of the other soliton (see Fig. \ref{fig1}). Considering the opposite frames of rotation in the CP solitons, the resonators are modulated with the added frequency components of the DKS combs. It is noteworthy to mention that using soliton crystals, it is possible to further shape the spectral distribution of the DKSs, which perturb the  resonators, thus improving the high-frequency components (or comb lines). Different moments of CP solitons mentioned above are shown in Fig. \ref{fig1}, to emphasize the interference and enhancement of the intensity of the intracavity field at moments when the solitons overlap perfectly. Although the synchronized resonators may have no perfect overlap (i.e., $M(\theta)=1, \, \forall \theta$), i.e., when we need them to have $T_{MW} = T_{opt}$, we consider such a case without loss of generality, as the modification toward the case of non-uniform overlap is straightforward by taking into account the non-uniform masking coefficient $M(\theta)$. This coefficient would only modify the Fourier-series coefficients describing the periodic modulation of the resonators. Therefore, in the rest of the paper, we consider the structure shown in Fig. \ref{fig1}, in which the optical and MW resonators are evanescently coupled with a constant coupling factor along the resonators. We also assume the MW-resonator has a fundamental frequency equal to $FSR_{opt}$ (i.e., $FSR_{MW} = FSR_{opt}$). Further discussions on the possible architectures for engineering the overlap between the optical and MW fields are discussed in Section \ref{sec:level4}.

\subsection{\label{sec:level21}Optical Microresonator}
Optical microresonators supporting coherent and stable solitons are demonstrated in a wide range of material platforms with reasonable static and dynamic control of their FSR. Example dynamic control mechanisms include the thermo-optic effect in controllable micro-heaters and  the Pockels effect. Adjusting the geometrical parameters of the microresonators (e.g., the microring radius), on the other hand, enables fixed resonators with desired FSRs.  This would imply possibility of adjusting the FSRs in the MW and optical resonators, which defines the frequency of the generated photons in the MW resonator. In this paper, we consider SiN microring resonators formed by bending a waveguide. To ensure the anomalous dispersion of the SiN waveguide (needed for bright soliton generation), we assume the optical structures are formed in a thick-film SiN platform \cite{48}, though more advanced dispersion engineering approaches based on coupled-resonators are possible in thin-film SiN platforms \cite{49, 50, 51, 52, 53}. The width and height of the SiN waveguide bent to make the microring resonator are chosen as $W_{SiN} = 1400 \, nm$, and $h_{SiN} = 800 \, nm$, respectively, to assure anomalous dispersion of the optical microresonator around the pumping wavelength of $\lambda_0  = 1550 \, nm$. The optical microresonator is assumed to have a radius of $R=200 \, \mu m$. Advances in the fabrication processes have shown promises for high-Q and dispersion-engineered SiN platforms, with $Qs > 10 \, M$ \cite{42, 43, 44}, yet we assume moderate Qs around 500 k, which is easily achieved using existing fabrication processes without high-temperature annealing or chemical mechanical polishing. The electrodes above and below the SiN layer, defining the MW resonators, have a width of $w=3.5 \, \mu m$ and are located at a distance of $1.5 \, \mu m$ away from the SiN boundaries.

\begin{figure}[t]
\centering
\includegraphics[trim = 120 200 100 50, width=0.5\linewidth]{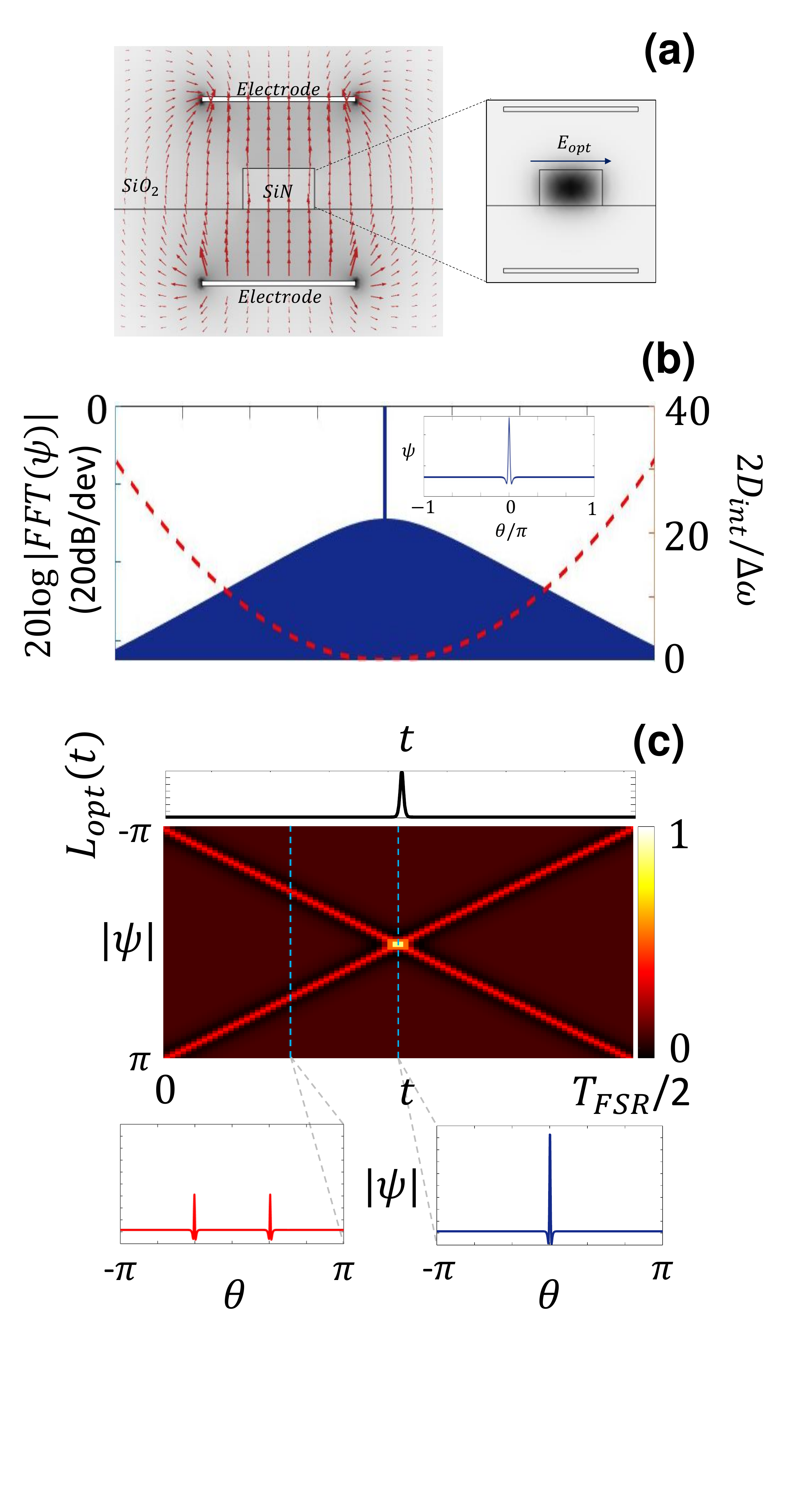}
\caption{\label{fig2} . Bright DKS generation inside the optical microresonator in the hybrid optical-MW configuration shown in Fig. \ref{fig1}, with the time-dependent spatio-temporal modulation of the hybrid MW and optical resonators. (a) Field profile of the MW resonator (red arrow) considering the plates above and beneath the SiN as PCs. Inset represents the intensity of the TE  field (i.e., electric field in the plane of the SiN waveguide) profile supported in the SiN waveguide buried in $\mathrm{SiO_2}$. The geometrical parameters of the SiN waveguide (width of $1400 \, nm$ and height of $800 \, nm$) are chosen to have anomalous dispersion around the wavelength $\lambda=1550 \, nm$. (b) The bright DKS soliton in a microring resonator with radius $R=200 \, \mu m$ and dispersion parameter $D_2=1.5246 \, MHz$, corresponding to the dispersing of the waveguide bent to form the microresonator. The dispersion in a structure with no strong curvature would follow that of the waveguide. The normalized power and detuning for generating the DKS are $F^2=4.1$, $\alpha=3.2$, respectively (inset: intracavity spatial distribution of the bright soliton). (c) The spatio-temporal distribution of the phase-locked CP solitons inside the microresonator mentioned in (b) by CP-CW pumps driving the microresonator. These CP solitons modulate the intracavity Kerr medium, perturbing the total optical path length with time (black inset), with periodicity $T_{opt}/2$ The intracavity signals for the CP solitons are also shown at two different moments: 1) when their spatial distributions perfectly overlap (blue curve) and when they do not (red curve). The electrodes with width $w=3.5 \, \mu m$ are at a distance of $1.5 \, \mu m$ away from the SiN boundaries.}
\end{figure}

The electric field profiles of the fundamental transverse electric (TE) mode in the SiN waveguide and the transverse electromagnetic (TEM) mode in the planar micro-strip (red arrow) are shown in Fig. \ref{fig2}(a). The Kerr-induced refractive index change in the optical microresonator, perturbs the MW-resonator, the strength of which is calculated using CMT through \cite{47}

\begin{equation}{\label{eq:CMT}}
\delta\beta^2=k_0^2 \frac{\int_S(n^2(x,y)-n_0^2)\psi_m^2 dS}{\int_S \psi_m^2 dS}
\end{equation}

with $n(x,y)$ and $n_0$ being the perturbed refractive index distribution (in x and y) through the Kerr effect and the effective refractive index with zero light intensity, respectively. In this relation, $\delta \beta$ is the perturbed propagation constant, $k_0$ is the magnitude of the free-space wavevector, and $\psi_m$ is the electric field distribution of the MW resonator in the transverse plane. The integrals in Eq. \ref{eq:CMT} are calculated over the surface in the transverse plane in Fig. \ref{fig2}(a). 

The Kerr-induced refractive index change in the SiN microresonator follows $\Delta{n}=\frac{\epsilon_0 n_0 n_2 c}{2}I$, with $I=|E_{cw}(\theta-\nu_g)+E_{ccw}(\theta+\nu_g)|^2$, and $E_{cw}$ and $E_{ccw}$ being the electric field distributions of the clockwise and  counterclockwise intracavity DKSs, respectively, and $\epsilon_0$, $n_2$, and $c$ representing the permittivity of the free space, nonlinear refractive index, and the speed of light in vacuum, respectively. It worth to mention that the masking coefficient ($M(\theta)$) modifies this Kerr-induced refractive index change, i.e., $n(\theta) \rightarrow n(\theta) \times M(\theta) $. This modification only affects the Fourier series coefficients expanding the periodic refractive index change through the Kerr effect. Thus, without loss of generality we consider the fixed masking coefficient $M(\theta)=1$. 

Figure \ref{fig2}(b) shows the spectral distribution of the DKS soliton in a microresonator with dispersion parameter $D_2=1.5246 \, MHz$, and normalized power and detuning $F^2=4.1$ and $\alpha = 3.1$, respectively. The loaded Q of $Q_l=500 \, k$ is assumed for the optical microresonator in this simulation. The spatio-temporal and spectral distribution of the Kerr-comb soliton is numerically studied using the split-step approach for implementing the Lugiato-Lefever equation (LLE) \cite{54}. The spectral distribution of the soliton is also indicated at the inset of Fig. \ref{fig2}(b). The spatial and temporal refractive index change in the optical microresonator supporting CP solitons are shown in Fig. \ref{fig2}(c), in the time window of $T_{opt}/2$, which is the periodicity of the refractive-index modulation. The total optical path length in the optical microresonator, defined as $\int_0 ^{2\pi} n(\theta,t) R d\theta$, is shown in Fig. \ref{fig2}(c), as well. There is a peak at the moment when the CP-DKSs perfectly overlap compared to the case when they do not overlap, as shown by the blue and red curves, respectively, in the inset of Fig. \ref{fig2}(c).

\subsection{\label{sec:level22}Effective Hamiltonian in MW-resonator}
The dynamical Casimir effect has been the subject of research in the last decade, with a range of analytical approaches to study the evolution of states under DCE. For a comprehensive review we refer the reader to Refs. \cite{55, 56, 57}. In this paper we follow the effective Hamiltonian approach to study the dynamics of the MW-resonator subject to the periodic perturbation caused by the CP solitons in the optical microresonator. For the rest of this paper, we assume fixed mode-overlap of $10\%$ between the MW-resonant mode in the transverse plane and the optical-intensity distribution shown in Fig. \ref{fig2}(a). 

The effective Hamiltonian for the dynamics of the system in Fig.\ref{fig1} (vacuum modes in the MW-resonator) is derived using the instantaneous eigenstates of the MW-resonator.

We define the instantaneous set of orthonormal eigenfunctions $\psi_n(\theta,t)$ with eigenfrequencies $\omega_n(t)$ satisfying the wave equation $\frac{\partial \psi_n (\theta,t)}{\partial \theta}=i \frac{\omega_n(t)}{c} n(\theta)R\psi_n(\theta,t)$ with normalization $\int_0 ^{2\pi} n^2(\theta,t) \psi_n(\theta,t) \psi_m(\theta,t) R d\theta=\delta_{nm}$ in which $\delta_{nm}$ is the Kronecker delta function, $n(\theta,t)$ is the intracavity spatial distribution of the refractive index, including the modification through the Kerr effect, $R$ is the radius, and $\omega_n(t)$ is the instantaneous angular frequency for the $n-{th}$ mode.

Quantizing the instantaneous eigenfunctions and introducing the time-dependent creation and annihilation operators ($a_n^+(t)$ and $a_n(t)$, respectively), we can write the intracavity field-operator $\phi(\theta,t)$ expanded in terms of the instantaneous eigenfunctions as $\phi(\theta,t)=\sum_n \frac{1}{\sqrt{2\omega_n(t)}}(a_n(t)\psi_n(\theta,t)+a_n^+(t)\psi_n(\theta,t))$. It is known that the annihilation operator $a_n (t)$ loses its meaning as the effective resonator length is changed with time, i.e., due to the change in the refractive index and the total optical path length \cite{58}. However, we can define an instantaneous annihilation operator for a fixed-length cavity and have a well-defined Hilbert space, at any specific time $t_0$, which is defined based on the instantaneous set of eigenfunctions. It should be noted that the annihilation operators defining vacua at different times $t$ are related to those at time $t=0$ via a unitary transformation (Bogoliubov transformation) \cite{59}. This linear transformation between bases at different times (through Bogoliubov transformation) makes them see the vacua of each other as squeezed states.

We study the effective Hamiltonian describing the dynamics of the system following the approach presented in Ref. \cite{58}. The effective Hamiltonian defining the evolution of the annihilation operator ($a_n(t)=U^+(t) a_n U(t)$) with a unitary operator $U(t)=Te^{-i\int_0^{2\pi}H_{eff}(t^\prime)dt^\prime}$, with $T$ being the time-ordering operator, and  $H_{eff}(t)$ being the effective Hamiltonian given the equation of motion $\frac{da_n(t)}{dt}=i[H_{eff},a_n(t)]$. The effective Hamiltonian reads as

\begin{multline}{\label{Heff}}
    H_{eff} =  i(a^+(t)C(t)a(t)+\\
    \frac{1}{2}a^+(t)D(t)a^*(t)-\frac{1}{2}a^T(t)D^*(t)a(t))
\end{multline}

In this relation, $a^*$, $a$ are column vectors consist of creation and annihilation operators for resonant modes in the MW-resonator, with their transpose $a^+$, and $a^T$, respectively.  In addition, time-dependent matrix elements in $C(t)$ and $D(t)$ are defined as
\begin{subequations}\label{eq:3}
\begin{align}
\label{eq:3a}
  &C_{nm}(t)=-i\omega_n(t)\delta_{nm}+\frac{1}{2}(G_{nm}(t)-G_{mn}(t)),\\
\label{eq:3b}
&D_{nm}(t)=\frac{1}{2}(\frac{\omega_n^\prime(t)}{\omega_n(t)}\delta_{nm}-G_{nm}(t)-G_{mn}(t)),
\end{align}
\end{subequations}
  
with

\begin{equation}
    G_{nm}(t) \equiv \sqrt{\frac{\omega_m(t)}{\omega_n(t)}}\int_0^{2\pi} n^2(\theta,t) \frac{\partial \psi_n(\theta, t)}{\partial t}\psi_m(\theta, t) R d\theta.
\end{equation}

To derive the effective Hamiltonian as a function of time, we evaluate the instantaneous set of eigenstates and eigenvalues, forming the orthonormal basis at each time, considering the effect of the Kerr-induced refractive index perturbing the MW-resonator. Moving to a reference frame generated through a unitary transformation $U(t) = e^{\sum_j i\omega_ja_j^+a_j}$, and using the Hadamard lemma to simplify the terms in Eq. \ref{Heff} along with the rotating wave approximation  to keep the modes resonantly excited, we can simplify the time-dependent $H_{eff}(t)$ to a more numerically convenient time-independent function. This is possible considering the periodic variations (with time) of elements in matrices $C(t)$, $D(t)$ in Eq. \ref{Heff}, which can be expanded in the form of Fourier series. To keep the resonant modes, it is useful to mention that term $a_k^+(t)a_l(t)$ varies with time as $e^{i(k-l)\omega_0t}$ in which $\omega_0$ is the fundamental angular frequency of the MW-resonator. This phase term is cancelled by its phase conjugate provided through the Fourier-series term of $C_{kl}(t)$ at angular frequency $(l-k)\omega_0$. In the same way, the terms $a_k^+(t)a_l^+(t)$ are in resonant with the $\mu-{th}$ term in the Fourier-series expansion of $D_{kl}(t)$, i.e., $D_{kl}^{\mu}$ for $\mu=-(k+l)$. This way, leveraging the periodicity and stability of the intracavity solitons, well-defined Fourier coefficients are extracted to numerically study the dynamics of vacuum states in the MW-resonator upon perturbation through the CP solitons in the optical microresonator.

Here we limit the numerical study to the Fock-space of three modes in the MW-resonator with each mode including up to $8$ identical photons. Although the average number of photons in each mode is much less than $8$ in the timeframe we considered to study the dynamics of the resonant modes in both ideal the MW-resonator with zero decay rate and the one with a finite decay rate, yet to avoid numerical errors we keep the higher number of photons. This will be reduced to smaller dimensions while trying to demonstrate the tomography of the density operator. It is trivial to adapt the same procedure for higher number of resonant modes, increased number of photons in each mode, and longer duration of the study.

\begin{figure}[t]
\centering
\includegraphics[trim = 140 10 280 50, width=0.6\linewidth]{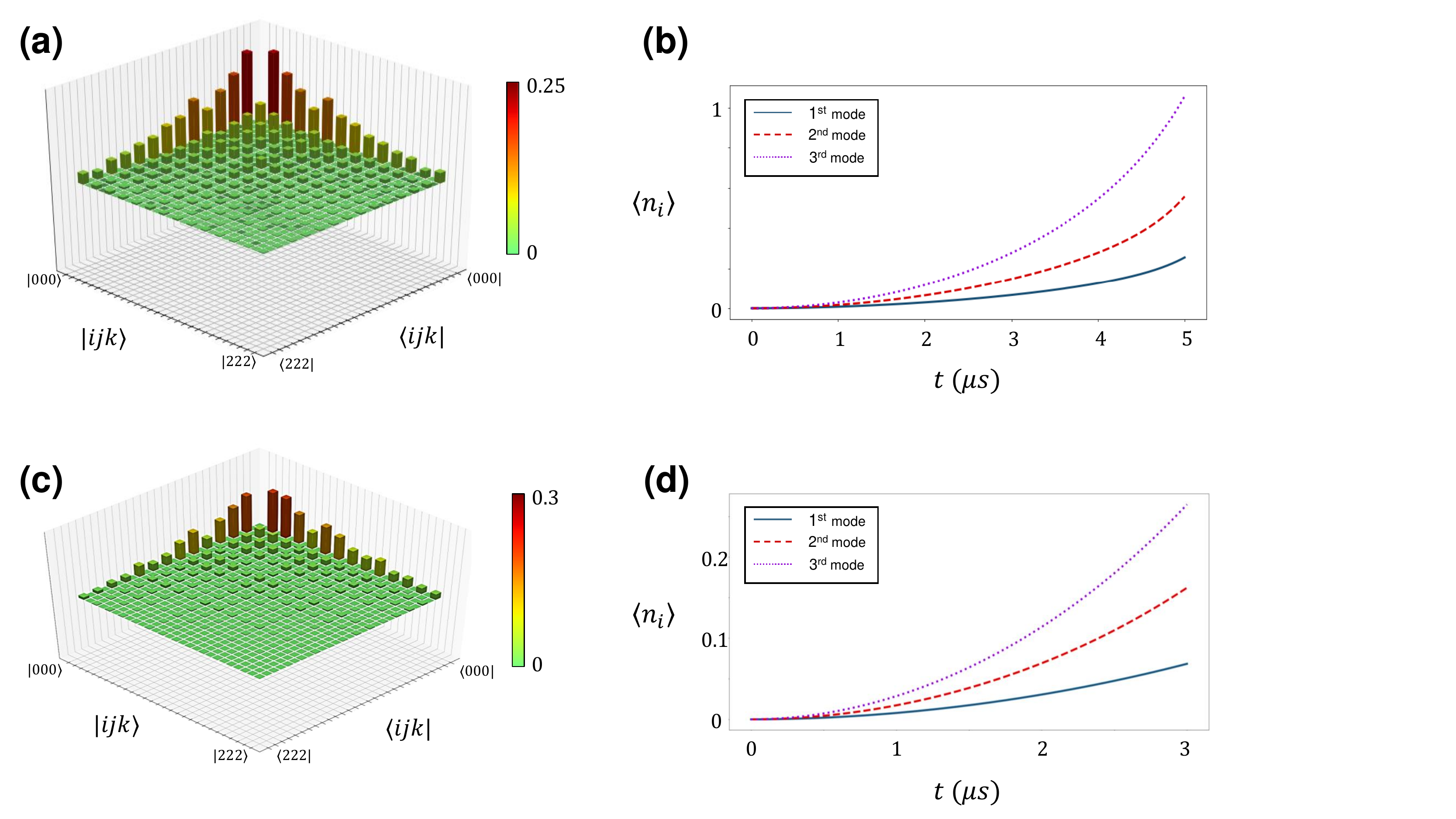}
\caption{\label{fig3} Multi-partite high-dimensional entangled photon state generation in a MW-resonator coupled to an optical microresonator supporting CP solitons with a fixed coupling factor $M(\theta)=1, \forall \theta$. The fundamental resonance of the MW-resonator matches the FSR of the Kerr-comb soliton. The MW-resonators with (a) zero decay rate and (c) a finite decay rate are studied. The former ensures a unitary operator and a pure output state, while the latter results a mixed state, making it complicated to study entanglement. In both cases, the mode overlap between the MW and optical resonators are assumed to be $10\%$, compatible with the field profiles shown in Fig. \ref{fig2}(a). In both simulations, three MW-resonant modes, each including up to $8$ photons are considered to ensure the accuracy of the numerical studies, yet for demonstrations, a subset of the density operator containing up to $3$ photons in each mode is considered. The CP solitons are assumed locked and identical to the one studied in Fig. \ref{fig2}(b). (a) State tomography of the density matrix for the truncated Fock-space of the MW-resonator with $3$ modes and a zero decay rate. The simulation is conducted in the time interval $[0-5\, \mu s]$. (b) Average photon population in each mode (blue solid: first, red-dashed: second, and dotted magenta: third) in the MW-resonator in (a). (c) State tomography of the density matrix in the MW-resonator with a limited decay rate of $10^5 \,  s^(-1)$ , corresponding to $Q = 1M$ for the optical microresonator. (d) Averaged photon number in the MW-resonator in (c). The dynamics in the time interval $[0-3 \, \mu s]$ is studied.}
\end{figure}

The simulations for studying the soliton formation in an optical microresonator (using the split-step method) are implemented in MATLAB, and those for the dynamics of the effective Hamiltonian are implemented in qutip-python \cite{60}. The system is studied in both cases of zero and finite decay rates for the MW-resonator. To demonstrate the photon generation through DCE, we consider both cases to take into account the finite decay rate of the MW-resonator, while for the study of entanglement we focus on the zero-decay-rate case. To study the effect of the cavity decay rate, we use the Lindblad master equation \cite{58} given by

\begin{multline}
    \dot{\rho}=-\frac{i}{\hbar}[H(t),\rho(t)]+\\
    \sum_n \frac{1}{2}[2C_n\rho(t)C_n^+-\rho(t)C_n^+C_n-C_n^+C_n\rho(t)],
\end{multline}

with $C_n = \sqrt{\gamma_n}A_n$ being collapse operators and $A_n$ being the operators modelling system coupling to the environment with corresponding decay rates $\gamma_n$. Further discussions about the the selected decay rate and corresponding resonator Qs can be found in Section \ref{sec:level4}.

Figures \ref{fig3}(a) and \ref{fig3}(c) show the tomography of the density operator for the MW-resonator states (limited to mode numbers $0-2$) with zero-decay and finite-decay rates, respectively. The average photon number in the three modes for these two cases are shown in Figs. \ref{fig3}(b) and \ref{fig3}(d), respectively. For the case with a zero decay rate, the evolution of MW-resonator vacuum modes under the effective Hamiltonian for $t=0-5 \, \mu s$ is studied, while for the finite-decay-rate case (with an identical decay rate of $\gamma=10^5 \, s^{-1}$ for all three modes) the evolution of these modes is studied for $t=0-3 \, \mu s$. The density matrix is shown in a subset of the total space, i.e., selecting photon numbers up to $3$ out of the total $8$ photons in each resonant mode. From these figures, we can see that a mixture of number states are generated thanks to the resonant excitation of resonator modes, in addition to the inter-mode coupling through the temporal and non-homogeneous spatial modulation of the optical and MW resonators by CP solitons.

\section{\label{sec:level3} Entanglement}

As previously mentioned, generating and benchmarking multi-partite high-dimensional entangled states are of particular importance in exploiting the quantum protocols and systems to suppressing the errors due limited coherence of qubits. To suppress the fragility of qubits, i.e., limited coherence time due coupling to the thermal reservoir, noises, etc., it is necessary to encode information on much robust states in the form of entangled physical qubits. It is shown that by entangling specific numbers of physical qubits, e.g., $5$ to $9$ qubits, it is possible to correct for error occurring in one of the qubits using error-correction protocols \cite{12}. This is a necessary step toward fault-tolerant quantum computing and an essential resource for other quantum protocols, e.g., teleportation and ultra-precise measurements. With the increased number of constituent parties forming entangled states, it is necessary to have an appropriate benchmark to measure the degree of entanglement. There have been several entanglement measures, e.g., entanglement of formation \cite{63}, relative entropy of entanglement \cite{64, 65}, negativity \cite{66,65}, and concurrence \cite{67, 65}, to name some of the most adapted ones; for further information please see Ref. \cite{65}. Among these measures, concurrence enables benchmarking multi-partite systems with different numbers of parties. If an $N$-partite state is bi-separable to an $(N-1)$-partite system and a one-partite state, the concurrence for the $N$-partite state reduces to that of the $(N-1)$-partite state.  In addition, concurrence automatically reduces to zero for fully separable states, and also it does not increase under LOCC; thus, it provides an entanglement monotone.

\begin{figure}[t]
\centering
\includegraphics[trim = 170 70 180 0, width=0.7\linewidth]{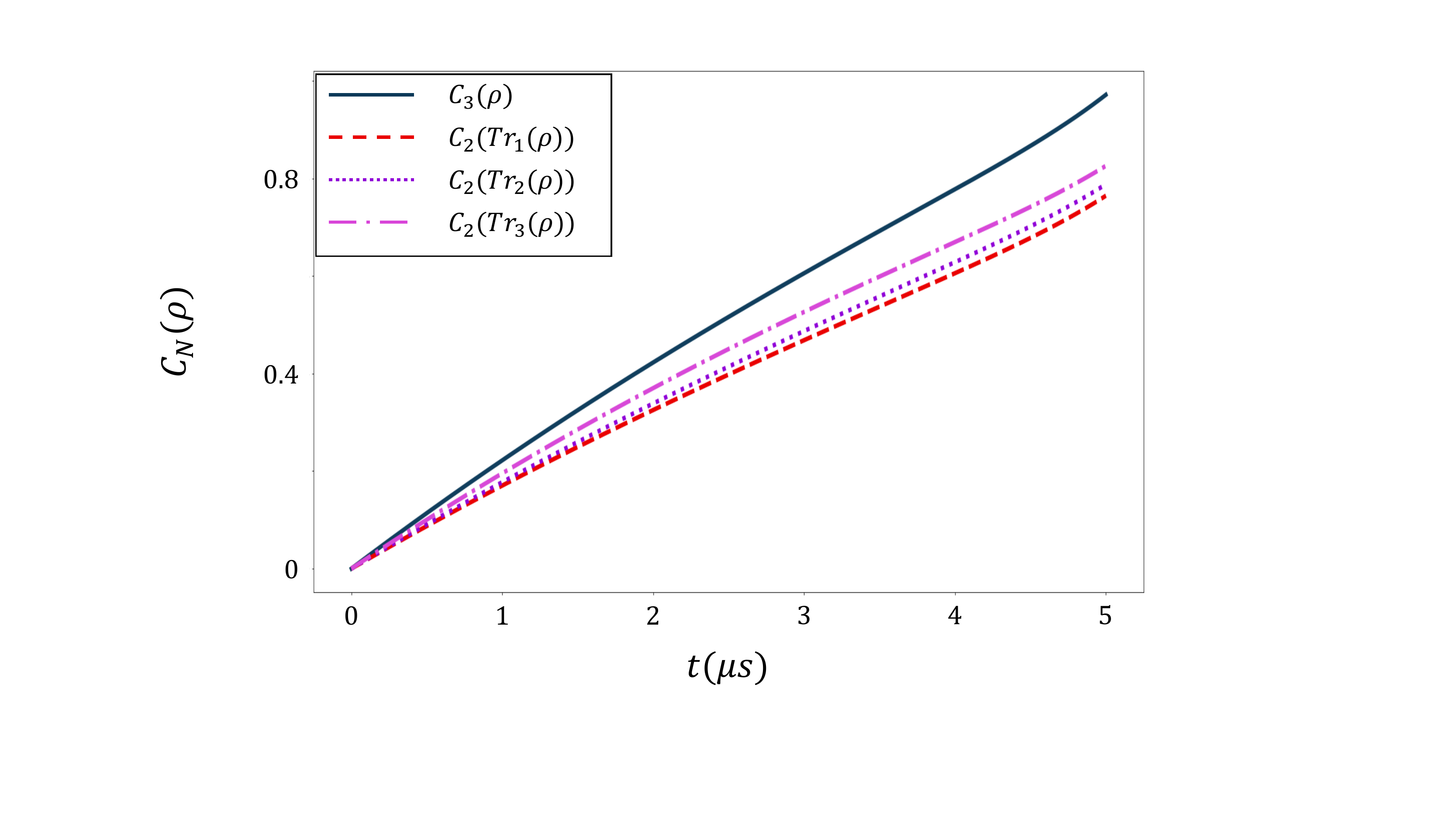}
\caption{\label{fig4} Concurrence of the pure state resulting from the effective Hamiltonian in the MW-resonator driven by CP solitons in the optical microresonator coupled to the MW-resonator (see Fig. \ref{fig1}). The solid blue line represents the concurrence of the three-partite state, while the dashed-red, dotted-magenta, and dashed-dotted light-magenta represent the concurrences of the reduced density matrices by tracing over one of the constituents.}
\end{figure}

In this section, we study the entanglement of the state in the MW-resonator with zero-decay (resulting pure state) as discussed in Section \ref{sec:level2} and shown in Fig. \ref{fig3}(a). The concurrence of an N-partite system is given by \cite{66}
   \begin{equation}
       C_N(\Psi_n) = 2^{1-N/2}\sqrt{(2^N-2)-\sum_iTr \rho_i^2},
   \end{equation}

where the index $i$ spans over all the $2^N-2$ subsets of the $N$-partite system, with $\rho_i$ being the reduced density matrix of the $1$ to $(N-1)$ multipartite subsystems, and $Tr$ being the trace operator. In the cases under study in this paper we evaluate the concurrence measure of the full (reduced) system with $N=3(2)$ to compare the concurrences of the two cases. For bi-separable states, the concurrence of the three-partite system is identical to that of one of the reduced two-partite states. In addition, we will study the concurrence of the collapsed density matrix upon measuring a photon in a specific mode (covering both cases of being able or not to distinguish the number states in the measured mode). Figure \ref{fig4} demonstrates the concurrence for the pure state (solid blue line) and reduced density matrices by partially tracing over one of the constituent parties (and reducing the space to a two-partite system). As seen from Fig. \ref{fig4}, the concurrence of the three-partite system is an increasing function of time, starting from zero, which corresponds to the fully separable vacuum states of the MW-resonator modes. It is also clear that the concurrence of the three-partite state is larger than those of all two-partite states. This indicates the state is not bi-separable. In addition, the reduced states, formed by partially tracing one of the parties, have non-zero concurrence, which highlights the entanglement of the reduced density matrix.

\begin{table*}[t]
\caption{\label{tab:table1}Concurrence of the collapsed state of the MW-resonator with zero-decay, evolved from vacua through the DCE of CP solitons, after $5 \, \mu s$, while measuring specific photon-number state in the qubit $\ket{i}$.}
\begin{ruledtabular}
\begin{tabular}{cccccccccc}
 
 $\ket{i}$ & $n=0$ & $n=1$ & $n=2$ & $n=3$ & $n=4$ & $n=5$ & $n=6$ & $n=7$ & $n=8$\\ \hline
 $i=0$ & 0.641 & 1.039 & 1.072 & 1.107 & 1.102 & 1.09 & 1.08 & 1.08 & 1.152 \\
 $i=1$ & 0.486 & 0.954 & 0.895 & 1.02 & 1.012 & 1.042 & 1.048 & 1.05 & 1.056 \\
 $i=3$ & 0.225 & 0.649 & 0.491 & 0.704 & 0.674 & 0.729 & 0.737 & 0.724 & 0.712 \\
 
\end{tabular}
\end{ruledtabular}
\end{table*}

\begin{figure*}[ht]
\centering
\includegraphics[trim = 0 0 0 0, width=0.9\linewidth]{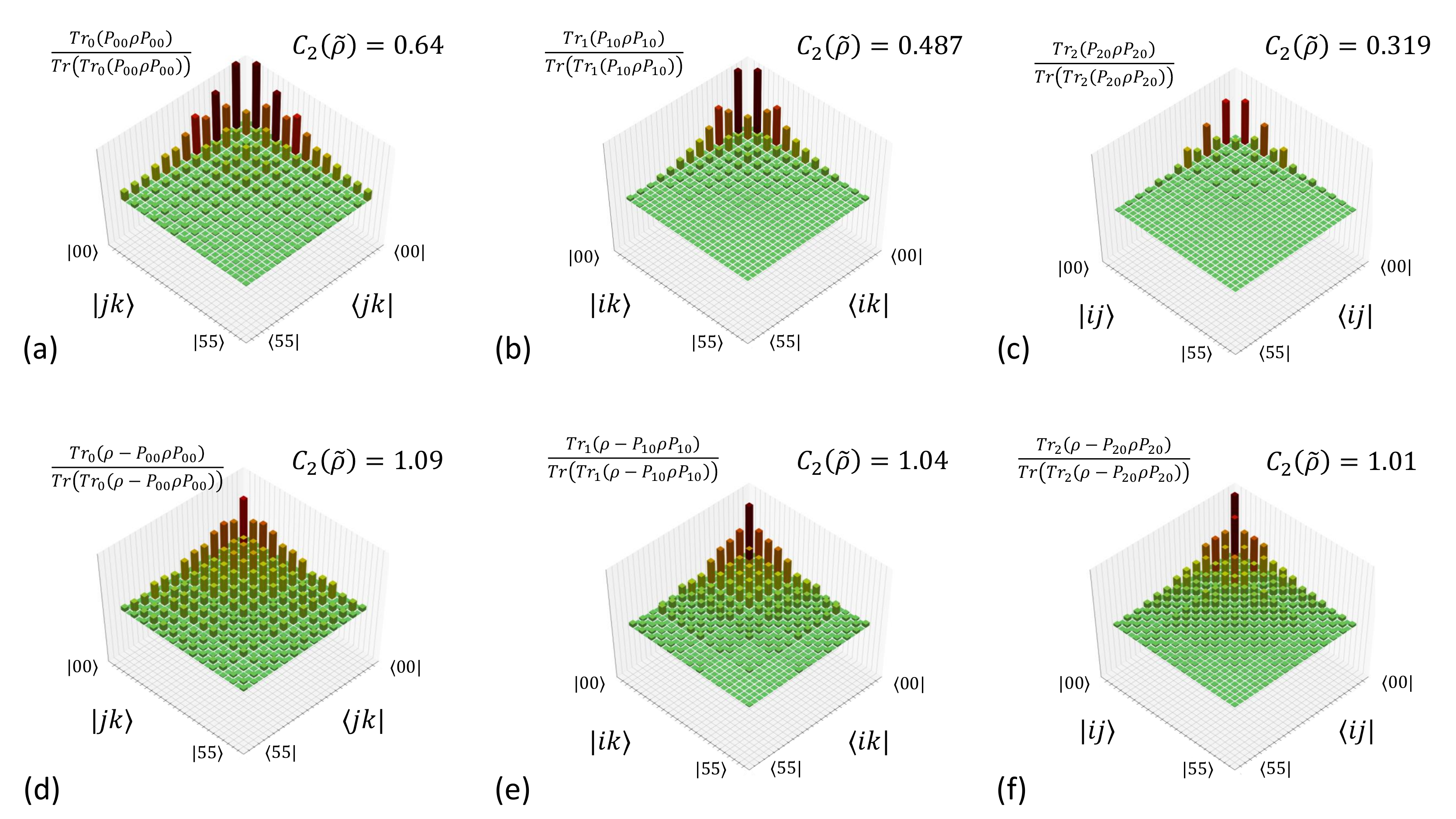}
\caption{\label{fig5} Projected states upon detecting  photons in each of the resonant modes (d), (e), (f) or measuring zero-photons at the modes (a), (b), (c) for the zeroth, first, and second-order modes, respectively, of the coupled resonator structure in Fig. \ref{fig1}. All the dimensions of the structure are the same as those in the Captions of Fig. \ref{fig2}. Operator $P_{ij}$ projects the state of the $i-th$ mode to specific mode number state $\ket{j}$.}
\end{figure*}

Figure \ref{fig5} shows the persistency of the entanglement upon detecting photons at a specific frequency (i.e., detecting or not detecting a photon at each of the three modes). A similar case for detecting a specific photon number in each mode is provided in Table \ref{tab:table1}. While the former can be easily implemented using single-photon detectors, the latter requires a more delicate detection scheme, e.g., photon number resolving detector \cite{73}, using strong coupling between a few photons and an ensemble of atoms by measuring the Rabi oscillation manifesting itself in characterizing the strongly coupled atom-photon system using a probe beam \cite{74}. Another option is to use a set of cavities to prepare the state that interacts with the target resonant mode and read the state, measuring the phase shift associated with the photon-number in the cavity mode \cite{QNDM}. Figure \ref{fig5} shows the case for the simple measurement on a resonant mode by detecting photon(s) (regardless of the photon numbers) or not detecting a photon. In the latter case, the state collapses to the projection of the pure-state onto the zero-photon at the corresponding mode, while the former is the residual orthonormal state by deducting the zero-photon projection. The tomography and two-partite concurrence of the reduced density matrix in each case is shown in Fig. \ref{fig5}, highlighting the entanglement persistency and multi-partite nature of the collapsed state. For further analysis, we also study the two-partite concurrence of the collapsed state upon measuring the specific photon-number state at each mode, summarized in Table \ref{tab:table1}. Again, from Table \ref{tab:table1} it is clear that the reduced two-partite system upon measurement is an entangled state (nonzero concurrence).

The results in Fig. \ref{fig5} clearly show that our coupled-resonator approach (see Fig. \ref{fig1}) is capable of generating high-persistency multi-partite high-dimensional entangled states. The collapsed density matrix upon measurement, and the associated concurrence measures in both Fig. \ref{fig5}, and Table \ref{tab:table1} clearly prove the persistency of the generated entangled states. Thus, our approach, paves the way toward highly persistent multi-partite high-dimensional entangled states for fault tolerant quantum protocols.

\section{\label{sec:level4} Fabricatoin Considerations}

Microelectronic fabrication processes developed and optimized for integrated photonic circuitry enable high-quality and low-loss platforms in a range of materials, e.g., SiN. During the studies presented in previous sections, we assume MW-resonator with internal Qs around $1 \, M$. To comment about the feasibility of such a platform, we need to briefly mention the progress on planar superconducting resonators with $Q > 1 \, M$, suitable for integration with integrated photonic structures \cite{67}. Considering the energy gap of superconductors, which implies the operational frequency on the order of a few GHz, such structures would be compatible with the current demonstrations of DKSs with FSRs of a few GHz \cite{37}. It is important to note that 3D SC cavities have much higher Qs \cite{68}, almost two orders of magnitude more than the planar structures, which may be well-suited for integration with the bulky and crystalline optical resonators (e.g., whispering-gallery-mode resonators in crystalline fluoride-based materials) \cite{69}. In addition, hybrid multi-layer structures are proposed to further increase the MW-resonator Qs \cite{70}. Better fabrication processes to ensure the crystalline structure of the SC films along with appropriate cleaning processes to diminish the unwanted residual impurities and defects at the surface of the SC indicate a path toward increasing the SC-resonator Qs \cite{67}. To further increase the operational frequency, it is necessary to have descent mode overlap or high-Q resonators. One option to increase the mode overlap could be leveraging high-Q photonic heterostructures with embedded natural two-dimensional electron gases (2DEGs) at the interface of III-V materials like AlGaAs or AlGaN. However, the Q of the MW-resonator formed by a 2DEG may not be large enough to allow the generation of photons. In addition, the high-energy optical photons would cause a high-temperature electron distribution, generating heat through scattering. There have been proposals on inducing superconductivity in 2DEG layers using short THz pulses, yet further developments are required to demonstrate high-Q resonators at the THz regime. \cite{71}.

\section{\label{sec:level5} Conclusion}

In this paper we demonstrated a fully integrated platform to generate multipartite high-dimensional entangled states through the dynamical Casimir effect in a MW-resonator subject to periodic perturbations through the CP solitons supported in an optical microresonator with the same FSR as that of the coupled MW-resonator. We used an effective Hamiltonian for the MW-resonator system based on instantaneous eigenstates to model the dynamics of the vacuum states in the MW-resonator with zero and nonzero decay rates. The numerical studies indicate the multi-partite entangled-state formation in the MW-resonator. To benchmark the degree of entanglement, we used a proper concurrence measure and studied the resulting pure state in a structure with zero decay rate for the MW-resonator. Through studying the persistency of the multi-partite entangled state, we proved the entanglement of the resulting collapsed state upon measurement on one of the resonant modes. We believe this system can provide a scalable high-dimensional multi-partite entangled source for use in a wide variety of quantum protocols, including but not limited to teleportation based on a fully integrated platform, beyond the three-partite system we studied in this paper.

\begin{acknowledgments}

We would like to thank Prof. Brian Kennedy from Georgia Institute of Technology for fruitful discussions on the entanglement measures and reviewing our paper. We would like to thank Prof. Victor Dodonov from University of Brasilia for helpful discussions and reviewing our DCE model. We would like to thank Ali A. Eftekhar for helpful discussions. Ali E. Dorche was partially supported through TI:GER (Technology
Innovation: Generating Economic Results) scholarship from Georgia Institute of Technology.
\end{acknowledgments}

\appendix

\nocite{*}

\bibliography{apssamp}

\end{document}